\documentclass[10pt,journal,compsoc]{IEEEtran}

\usepackage{amsmath,mathtools,amsfonts,amssymb,inputenc,graphicx,comment,tikz,tikz-cd,soul,float,algorithm,algpseudocode}

\makeatletter
\algnewcommand{\LineComment}[2][0]{\Statex
  \ifnum0=#1 \hspace*{\ALG@thistlm}%
  \else\hspace*{\ALG@tlm}%
  \fi
\(\triangleright\) #2}
\newlength{\CAS@tlm}
\newlength{\CAS@triIndent}
\algnewcommand{\ParComment}[2][0]{%
\ifnum0=#1\setlength\CAS@tlm{\ALG@thistlm}%
\else\setlength\CAS@tlm{\ALG@tlm}%
\fi%
\Statex%
\hspace*{\CAS@tlm}
\parbox[t]{\dimexpr\linewidth-\CAS@tlm-3pt}{\hangindent=\CAS@triIndent\strut \( \triangleright \) #2\strut}}
\algnewcommand{\ParState}[1]{%
\State \parbox[t]{\dimexpr\linewidth-\ALG@thistlm}{\strut #1\strut}%
}
\newlength{\CAS@forallwidth}
\newlength{\CAS@forwidth}
\newlength{\CAS@ifwidth}
\algdef{S}[FOR]{ParFor}[1]
  {\parbox[t]{\dimexpr\linewidth-\ALG@thistlm}{%
     \hangindent\CAS@forwidth\strut\algorithmicfor\ #1\ \algorithmicdo\strut}}%
\algdef{S}[FOR]{ParForAll}[1]
  {\parbox[t]{\dimexpr\linewidth-\ALG@thistlm}{%
     \hangindent\CAS@forallwidth\strut\algorithmicforall\ #1\ \algorithmicdo\strut}}%
\algdef{S}[IF]{ParIf}[1]
  {\parbox[t]{\dimexpr\linewidth-\ALG@thistlm}{%
     \hangindent\CAS@ifwidth\strut\algorithmicif\ #1\ \algorithmicthen\strut}}%
\makeatother

\newcommand{\procSep}{\vspace{0.75em}}

\usepackage{hyperref}

\newcommand{\defeq}{\mathrel{\mathop:}=}
\renewcommand{\and}{\wedge}

\RequirePackage{amsthm}
\newtheorem{theorem}{Theorem}[section]
\newtheorem{lemma}[theorem]{Lemma}

\usepackage[nocompress]{cite}

\title{Casanova}

\author{Kyle~Butt,
        Derek~Sorensen,
        and~Michael~Stay
\IEEEcompsocitemizethanks{\IEEEcompsocthanksitem
K. Butt (kyle@pyrofex.net) is a Staff Software Engineer at Pyrofex Corporation.
\IEEEcompsocthanksitem
D. Sorensen (derek@pyrofex.net) is a Research Mathematician at Pyrofex Corporation.
\IEEEcompsocthanksitem
M. Stay (stay@pyrofex.net) is CTO of Pyrofex Corporation.
}
\thanks{Manuscript received March 01, 2019; revised March 31, 2019.}}

\begin{document}

\IEEEtitleabstractindextext{%
\begin{abstract}
We introduce Casanova, a leaderless optimistic consensus protocol designed for a permissioned blockchain. Casanova produces blocks in a DAG rather than a chain, and combines voting rounds with block production by singling out individual conflicting transactions.
\end{abstract}

}

\maketitle
\IEEEdisplaynontitleabstractindextext
\IEEEpeerreviewmaketitle

\section{Introduction}

The problem of achieving consensus over an asynchronous network between mutually distrustful parties has been a subject of research for several decades. After Nakamoto introduced Bitcoin \cite{BitcoinWhitePaper}{,} a new class of consensus algorithms emerged, making use of either miners or some kind of leader election. These include: proof of work (PoW), proof of stake (PoS), and proof of elapsed time (PoET) protocols, among others. Classical protocols, by contrast, tend to rely on rounds of message passing\cite{PaLa}{.}
Proof of work algorithms are inherently inefficient because their security properties come from ``wasted" computation, \textit{e.g.} brute forcing hash preimages. The blockchain community is naturally interested in finding reliable and scalable consensus algorithms suited to use in a proof of stake context.



We present \href{https://www.youtube.com/watch?v=JuhTQwzizDI}{Casanova}, a leaderless optimistic consensus protocol optimized for the blockchain, but which borrows from pre-Nakamoto consensus protocols. It would be appropriate for a proof-of-stake blockchain, but can be used in a variety of applications. 
It shows that one can pipeline voting and message-passing rounds by combining block creation with votes, and optimizes for the overwhelmingly common case that a transaction is \textit{not} a double spend. 

Casanova emerged from two principal observations. The first is that, as a rule, transactions on a blockchain do not conflict. The majority of users do not double spend because they want their transactions to clear quickly. Despite this fact, modern consensus protocols spend resources coming to consensus among alternatives on \textit{every} transaction, regardless of whether or not any conflicting transactions have been seen. This is excessive and inefficient. Casanova achieves safety, liveness, and scalability by primarily running a \textit{conflict exclusion protocol}, using a choice consensus protocol only when necessary for the occasional double spend, and only for the conflicting transactions. Conflicting transactions do not delay non-conflicting transactions in the same block.

Classical consensus algorithms are generally not suited to take advantage of this apparent lack of conflict because they make two assumptions that are not true in a blockchain context:
\begin{enumerate}
    \item Processors begin with an initial value $v \in V$, where $V$ is the value domain.
    \item Values $v \in V$ are readily forged. This assumption is usually implicit in the defenses against Byzantine behavior.
\end{enumerate}
Casanova is able to take advantage of the general absence of double spends by removing these assumptions.

The second observation is that transactions do not need to be totally ordered. A partial order will suffice, as most transactions are unrelated. Current blockchains record transactions in a single chain, creating a total order between all transactions. Unrelated transactions do not need to be ordered, only related transactions (\textit{e.g.} I received some money I later spent). By structuring transactions in a partial order, rather than a total order, Casanova is able to allow each validator to produce blocks at a regular interval rather than requiring mining or leader election to produce a single block at a regular interval.

For this reason, in Casanova, blocks do not form a chain, but a \textit{directed acyclic graph}.

\section{Preliminaries}

\subsection{Network Requirements}
To achieve liveness we assume \textit{partial synchrony}, which is that the network delivers any message within some finite time bound $\Delta$. We do not need to know, or even have an estimate of $\Delta$ at any point. To achieve safety, on the other hand, our network can be totally asynchronous. Messages can be arbitrarily dropped, reordered, or duplicated.

\subsection{Events and Transactions}
We assume some set of all possible events, $\mathbb{E}$. We also assume that $\mathbb{E}$ is suitably partitioned into mutually exclusive alternatives:
$$\mathbb{E} = \bigsqcup\limits_{i \in I} \mathcal{E}_{i}$$
for some indexing set $I$. That is, each event $e \in \mathbb{E}$ belongs to exactly one $\mathcal{E}_{i}$ and at most one such event could have occurred.
We call the events \textit{transactions} if there exists some $\mathcal{E}_i$ that contains more than one event.
We assume from any transaction $e_{i}$ that the conflict index $i$ can be computed, and that we have a suitable function $index: index(e_{i} \in \mathcal{E}_{i}) = i$

\subsection{Graph theoretic notions}

A \textit{directed acyclic graph} (DAG) is a directed graph with no cycles. A cycle is a path along graph edges whose source and target are the same vertex. In the case of Casanova, any edge in the DAG has as its source a child block and as its target a direct parent of the child. 
A \textit{leaf} is a graph vertex that is the source to some (possibly empty) set of edges but target to none.

Note that a finite nonempty DAG always has a nonempty set of leaves. To find a leaf, one can choose a vertex at random and follow any reverse path until termination. In the DAG consisting of one vertex (and no edges), that vertex is a leaf. In particular, the genesis block alone is a leaf. To disambiguate nodes in a DAG from nodes in the network, we use the term \textit{validator} to refer to a network node, and \textit{block} to refer to a node in the DAG.

Casanova builds a DAG of blocks instead of a chain. Each block contains a reference to one or more parents. An individual validator is not allowed to produce a block that is not a descendant of their most recent block. Violating this requirement is called \textit{equivocation} and is a fault.

We will occasionally abuse notation and treat a DAG as either a set of blocks or a set of transactions. When we treat it as a set of blocks, we mean the underlying set of blocks, and when we treat it as a set of transactions, we mean the union of transactions over the underlying set of blocks. We also use $+$ to represent adding a block to a DAG. The meaning should be clear as the block contains its parent links.

\subsection{Consensus on a DAG}
A DAG-consensus algorithm for a set of partitioned events $\mathbb{E} = \bigsqcup\limits_{i \in I} \mathcal{E}_{i}$ is a protocol together with a family of functions $decide_{i}$, $i \in I$, from DAGs to $\mathcal{E}_{i} + \mathcal{P}(\mathcal{E}_{i})$, the disjoint union of $\mathcal{E}_{i}$ with the powerset of $\mathcal{E}_{i}$. It either returns the decision or a set of known alternatives when no decision has been reached.
If $decide_{i}(D) \in \mathcal{E}_{i}$, then we say a validator with DAG $D$ has decided on $e_{i} = decide_{i}(D)$.
If $decide_{i}(D) = \varnothing$, then $D$ does not include any transaction $e_{i} \in \mathcal{E}_{i}$.
If $decide_{i}(D) = E_{i} \in \mathcal{P}(\mathcal{E}_{i}) \and E_{i} \neq \varnothing$, then $E_{i}$ is the set of \textit{currently known} alternatives. In particular if $\vert E_{i} \vert = 1$, this is not the same as a decision. In the absence of new alternatives a representative from $E_{i}$ will eventually be chosen.

Stating the requirement about ``eventually chosen'' more formally:
If $decide_{i}(D) = E_{i} \subset \mathcal{E}_{i}$, ${E_{i} \neq \varnothing}, {D \subset D'}, decide_{i}(D') \in \mathcal{E}_{i}$ and $D'$ does not contain any transactions in $\mathcal{E}_{i}$ not already in $D$, then $decide_{i}(D')$ must be an element of $E_{i}$.

\subsubsection{Safety}
Assume that we have a DAG $D$ as seen by a validator $V$ such that $e_{i} = decide_{i}(D) \in \mathcal{E}_{i}$ for some $i$. For every validator $W \neq V$ let $D_{W}$ be the induced sub-DAG created by taking the most-recent blocks created by $W$ and every ancestor block of those blocks. A DAG-consensus algorithm is \textit{safe} if for every correct validator $W \neq V$ the protocol cannot create a DAG $D'$ which contains $D_{W}$ as a subgraph s.t. $e'_{i} = decide_{i}(D') \in \mathcal{E}_{i}$ and $e'_{i} \neq e_{i}$. More succinctly: $$\not\exists D'. D_{W} \subset D' \and e'_{i} = decide_{i}(D') \in \mathcal{E}_{i} \and e'_{i} \neq e_{i}.$$ Informally, this means that if a correct validator decides, no other correct validator can make a different decision.

\subsubsection{Liveness}
A DAG-consensus algorithm is \textit{live} if the protocol guarantees that for all $i \in I$, a correct validator $V$ either has only seen DAGs $D$ such that $decide_i(D) = \varnothing$ or $V$ 
eventually sees a DAG $D$ such that $decide_i(D) \in \mathcal{E}_{i}$.

\subsection{Algorithm Structure}
We use the communicating event loop model, in which a single thread pulls an event from a queue and processes it before handling the next event.  Each of the algorithms included is presented as an initial state and several event handlers. The event handlers each include a comment indicating the state that gets read. We underline any state that gets modified. Events may contain information; that information is presented as arguments to the event handler.

\subsection{Validators and Byzantine behavior}
A {\em Byzantine} validator is one which behaves arbitrarily. We denote by $N$ then number of validators and $f$ the number of Byzantine validators in our network. A \textit{correct} validator is one that is not byzantine. We define a \textit{Non-Faulty Majority} ($NFM$) as $\lceil\frac{N - f + 1}{2}\rceil$, and a \textit{Fault Tolerant Majority} ($FTM$) as $\lceil\frac{N + f + 1}{2}\rceil = NFM + f$. Given an $NFM$, any $FTM$ will contain at least one validator from the $NFM$.

\subsection{Layout of what follows}
The rest of the paper has the following structure: In \S\ref{Sctn: Casanova-attest} we build Casanova-\textit{attest}, a simple attestation protocol that gives structure to the trivial consensus protocol. In \S\ref{Sctn: Casanova-conflict-attest} we expand Casanova-\textit{attest} to Casanova-\textit{conflict-attest}, which can handle conflicting transactions by using an auxiliary consensus protocol. After defining some relevant terms in \S\ref{Sctn: DAG Properties}, in \S\ref{Sctn: Casanova-conflict-exclude} we expand Casanova-\textit{conflict-attest} to Casanova-\textit{conflict-exclude}, which shows when we can avoid having to use the auxiliary protocol. Finally, in \S\ref{Sctn: Casanova-final} we show how to build consensus natively into the structure with Casanova.

\newpage 
\section{Using attestation to identify and resolve conflicts}
\subsection{An attestation protocol} \label{Sctn: Casanova-attest}
In the majority of blockchain applications, the network's responsibility more closely resembles one of attestation rather than consensus. Here we use \textit{attest} to mean \textit{record with evidence}. More specifically, the problem of attestation is to record events supplied by users in a partial order. Users can request that a particular event $e$ follow some other event $e^{\prime}$. In the absence of double spends, this is exactly what a blockchain is doing, as creation of outputs must precede the spending of those outputs. We sketch a simple attestation protocol Casanova-\textit{attest}: 

Starting with the empty record (a genesis block), validators build a DAG both by adding blocks of their own and by listening to their peers. To produce its own block, a validator listens for \textit{(event-data, parent-hashes)} tuples to record from clients. The \textit{parent-hashes} are hashes of events that must have been recorded in order to record this tuple. If a validator has recorded events matching all the parent hashes the user requested, it will add the tuple to its next block. Otherwise it ignores the request. Validators periodically produce blocks $b$ whose events are all of the unrecorded tuples they have received and whose parents are all the leaves of the DAG they have seen so far. Once a validator adds $b$ to its DAG, it communicates $b$ to the rest of the validators. For pseudocode, see Algorithm \ref{attest}.

When a validator receives a block from a peer, if it has all of the block's parents, it adds the block to its DAG. If not, it waits until it has the parents in its DAG, and then adds the block to its DAG.



If a validator has recorded an event, nothing in the protocol allows it to alter this record. Assuming all messages are eventually delivered, every correct validator will eventually record the event as well. In essence, Casanova-\textit{attest} is a trivial consensus protocol---consensus over a set of size 1---where attestations grow in a DAG. Note that this does not contradict the standard impossibility result\cite{FLP}{,} because we are essentially building consensus on sets of size 1.

\subsection{From attestation to consensus} \label{Sctn: Casanova-conflict-attest}
Casanova-\textit{attest} only works if there is no possibility of conflict between the events to be recorded. We call events with the possibility of conflict \textit{transactions}. Assume that we can handle up to $f$ Byzantine faults, where $N \geq 3f + 1$. We modify Casanova-\textit{attest} as follows to produce Casanova-\textit{conflict-attest}:

When a validator receives two blocks that contain conflicting transactions $e_{i}, e'_{i} \in \mathcal{E}_{i}$, it attests to the fact that it observed the conflict. The validators can then use some other means (like Algorithm 2\textsuperscript{2} from ``Consensus in the Presence of Partial Synchrony'' \cite{DLS}) to come to consensus on some representative $e_{i} \in \mathcal{E}_{i}$. Upon achieving consensus, each validator records the decision and the evidence produced by the side protocol in the DAG. Assuming that some validator sees the conflict first and the others see it within $\Delta$, the side protocol will have a delay of at most $2\Delta$, and so will eventually decide. For pseudocode, see Algorithm \ref{conflict-attest}.

\subsection{Selective consensus}
If every transaction admits at least two conflicting alternatives, the above protocol could require a run of Algorithm 2\textsuperscript{2} for every transaction. 
It would be nice if we could lean on the attestation DAG to avoid some of these runs. We will need to introduce some DAG terminology in order to describe a protocol that does so. 

\section{DAG Properties} \label{Sctn: DAG Properties}

\subsection{Scoring blocks in a DAG}
We want to solve the problem of assigning a set of validators (voters) to the blocks in a DAG. We call this set the {\em score}. Given the set, we can compute either a count or, more generally, a weight. We are computing the score set as seen by validator $V$. The score of a block is the union over all self-or-descendant blocks of the validators that created those blocks. More formally, for each block $b$, let $W_{b}$ be the validator that created the block.  The score of $b$ as computed by $V$ is $$score_{V}(b) \defeq\{W_{b}\}\cup\bigcup\limits_{c \in children(b)} score_{V}(c).$$ 
This can be efficiently computed via any reverse topological traversal.

\subsection{Observed Sets}
Each validator $V$ also needs to compute the score as seen by other validators when they created their blocks. In $V$'s view of the DAG, there is one \textit{most recent} block created by each correct validator. If there is more than one such block for a validator $U$, $U$ must have equivocated. A block $b$ is most recent for validator $U$ if $V$ has not seen a block created by $U$ that is a descendant of $b$.
A validator's most recent block may be the genesis block, which is treated as if it were created by all the validators. For each validator $U$, $V$'s view of $U$'s score is computed as above, but on the induced subgraph of the most recent block. If a validator $U$ has two or more most-recent blocks, $V$ can simply exclude them entirely and reduce $N$ and $f$ both by one, as $V$ has proof that $U$ is faulty. We will represent $V$'s view of $U$'s score at block $b$ as $viewscore_{V}(U, b)$

Assume some weighting function $weight$ which maps validators to integer weights. In the simple case, $weight$ can be the constant function $1$. We can extend this function to sum over a set of validators. Given such a function, we define a \textit{$k$-observed set} as follows: A \textit{$k$-observed set}, for a block $b$ is a set $S$ of validators such that: $$\forall U \in S: weight(viewscore_{V}(U, b)) \geq k$$ and $$weight(S) \geq k.$$ There is a straightforward algorithm to check for the existence of a $k$-observed set.

We will compute the score of transactions in a similar manner, and unless otherwise modified, the score of a transaction is the score of the block that contained it. If a transaction is in more than one block, its score is the union of the scores of the blocks that contain it.

For the purpose of conflict resolution, some attestations have rounds attached to them. When looking for a $k$-observed set, the attestations must all belong to the same round. Blocks from later rounds should be excluded when traversing the DAG, and blocks from earlier rounds don't count when calculating the score for this round.

\section{Using the DAG to remove the side consensus protocol}
We now have enough DAG machinery to allow attestation to do some of the consensus work. 
\subsection{Selective Consensus: Exclusion} \label{Sctn: Casanova-conflict-exclude}
We change Casanova-\textit{conflict-attest} as follows to produce Casanova-\textit{conflict-exclude}. Note that the attestations described here only apply to finding an $FTM$-observed set, and are not part of the side protocol.

When a validator $V$ receives two blocks that contain conflicting transactions $e_{i}, e'_{i} \in \mathcal{E}_{i}$, it attests to the fact that it observed the conflict. It then computes the scores of the conflicting transactions. If any validator had attested to a transaction before attesting to the conflict, their attestation toward the first transaction is included in the score, including $V$ if it attested. If $V$'s observed score for the block containing $e_{i}$ is $FTM$ or greater, without seeing an $FTM$-observed set, $V$ joins the side consensus protocol with altered initial conditions. For example, if we use Algorithm 2\textsuperscript{2}, $V$ would join the protocol with a phase $-1$ lock on $e_{i}$. See the Appendix for a summary of Algorithm 2\textsuperscript{2}. There may be other consensus algorithms that admit suitably altered initial conditions.

If $V$ does not see a score of $FTM$ or greater for either $e_{i}$ or $e'_{i}$, it joins the side consensus protocol with standard initial conditions. If $V$ sees an $FTM$-observed set, it does not join the side consensus protocol, but rather continues to produce blocks attesting to the fact that it observed such a set. This gives us two cooperatively overlapping decision criteria: the first is observing an $FTM$-observed set, and the second is receiving the decision of the side consensus algorithm. For pseudocode, see Algorithm \ref{conflict-exclude}.


\begin{lemma}
If an $FTM$-observed set exists, then Algorithm 2\textsuperscript{2} can only decide on the value attested by the $FTM$-observed set.
\end{lemma}

The proof is a straightforward alteration of Lemma 3.5 from ``Consensus in the Presence of Partial Synchrony.'' \cite{DLS}{,} because at least $NFM$ correct processors start with a lock on $e_{i}$ at phase $-1$.
This means that if a validator receives evidence of an $FTM$-observed set, they can stop the side protocol, as the outcome is a foregone conclusion.

It is straightforward to see that if no such observed set exists, then the correct processors join the side protocol within $\Delta$, and so the side protocol has a delay of at most $2\Delta$, and will therefore converge.

\subsection{Completely removing the side consensus algorithm} \label{Sctn: Casanova-final}

We can now build in a consensus algorithm natively into Casanova-\textit{conflict-exclude}, and dispense with a side algorithm. We call this Casanova.

We assume that validators produce a new block every time unit $\delta$. A round is defined as a fixed number of blocks.
Round 1 for a validator begins two blocks after the block that begins round 0. Each round $r \geq 1$ begins $r + 1$ blocks after the block that began round $r - 1$ (that is, round $r$ starts $\frac{r(r + 3)}{2}$ blocks after round $0$). Validators can have locks on a value which is internal state not externally visible. Locks on a value have an associated round. Once a validator has a lock on a value, it may release it only to take a lock with a later associated round. This may be for the same or a different value.

The state before a conflict is encountered is treated as round $-1$. Upon encountering conflicting transactions $e_{i}$ and $e'_{i}$, round $0$ for the conflict domain $\mathcal{E}_{i}$ begins. If a validator has observed $e_{i}$ with a score of $FTM$ or greater, it has a round $-1$ lock on $e_{i}$. If a validator observes some transaction $e_{i}$ with a score of $FTM$ or greater in any round $r$, it takes a lock on $e_{i}$, with the associated round $r$. If it has a lock from a previous round $p < r$, it releases the earlier lock and takes the later lock.

The validators must vote at the beginning of each round $r$ according to the following rules:
\begin{enumerate}
\item If the validator has a lock on $e_{i}$ from a previous round $p < r$, the validator attests to their vote for $e_{i}$ at the beginning of round $r$, and the DAG includes their attestation that they did see a score of $FTM$ or greater in some previous round.
\item Otherwise, the validator records a vote for transaction $e'_{i} \in \mathcal{E}_{i}$ with the lowest hash seen so far. (Any other deterministic ordering would work as well)
\end{enumerate}
To bound the number of possibilities from a particular exclusion set $\mathcal{E}_{i}$, each validator may attest to at most 1 transaction from $\mathcal{E}_{i}$. Conflicts may arise due to different validators attesting to different transactions before seeing blocks containing conflicting transactions. This bounds the number of transactions that will be considered from any given $\mathcal{E}_{i}$ to $N - f + f(N - f)$.

If, for any $r$, a validator sees a round-$r$ $FTM$-observed set for a transaction $e_{i}$, the validator decides on $e_{i}$, and attests to that fact. The validator no longer counts rounds, but can continue with the attestation. Other validators will eventually also see the same decision criteria. For pseudocode, see Algorithm \ref{casanova}.

\subsection{Formal Decision Criteria}
We define $decide_{i}$ as follows for Casanova. $decide_{i}(D) = e_{i}$ if $e_{i} \in \mathcal{E}_{i}$ and $D$ contains an $FTM$-observed set having voted for $e_{i}$ in some round $r$. If an $FTM$-observed set exists in more than one round, we take the value chosen in the earliest round. Otherwise $decide_{i}(D) = \lbrace e \in D \vert index(e) = i \rbrace$.
Assuming we have $f$ or fewer Byzantine validators, this is well defined because the existence of two or more $FTM$-observed sets for different transactions $e_{i} \neq e'_{i}$ would require at least $f + 1$ equivocations.

\subsection{Proof of Safety for Casanova}
Assume some validator $V$ with DAG $D$ s.t. $e_{i} = decide_{i}(D) \in \mathcal{E}_{i}$, and let $r$ be the round associated with the decision. Then $V$ observed an $FTM$-observed set. Of the $FTM$-observed set, at least an $NFM$ must be locked on $e_{i}$. In order to unlock from $e_{i}$, they must observe a transaction $e'_{i} \in \mathcal{E}_{i}$ in round $r' > r$ with score $FTM$ or greater. But because at least $NFM$ are locked on $e_{i}$, the greatest possible score that could be seen is $FTM - 1$. This precludes the formation of an $FTM$-observed set on $e'_{i} \neq e_{i}$, therefore no two correct processors can decide on different values under the assumed fault bounds. There is some earliest round $r_{min}$ where any validator decides. The above shows that if any correct validator decides a round $r > r_{min}$, they will make the same decision.
\hfill $\square$

\subsection{Proof of Liveness for Casanova}
We will prove liveness for some particular $\mathcal{E}_{i} \subset \mathbb{E}$; specifically, at least one representative $e_{i} \in \mathcal{E}_{i}$ has been observed by at least one correct validator. Assume a validator $V$ with DAG $D$ s.t. $decide_{i}(D) \in \mathcal{P}(\mathcal{E}_{i}) \and decide_{i}(D) \neq \varnothing.$
Assume that round $r$ is longer than $2 \Delta$. Then before the start of round $r+1$, there is at most one locked transaction: during the round, all validators receive evidence of locks from all correct validators. If two or more locks exist, the correct validators will release all but one of them, the lock with the highest round. If such a lock exists, then every correct validator has received evidence for it. Then the blocks that start round $r + 1$ will be received in enough time for an $FTM$-observed set to form during the round. If a lock does not exist at the start of round $r + 1$, a lock will occur within $N - f + f(N - f)$ additional rounds, because of the bound on the number of possible conflicts that may be accepted.
\hfill $\square$

\begin{algorithm}[ht]
\caption{Casanova-\textit{attest}}\label{attest}
\begin{algorithmic}[1]
\State $dag \gets (\lbrace genesis \rbrace, \varnothing)$
\State $pending, waiting \gets \varnothing$
\ParComment{Symbols underlined in the "Uses:" comments may be modified in the corresponding procedure}
\Procedure{ReceiveBlock}{$b$}
    \LineComment[1]{Uses: $\underline{dag},\underline{pending}$}
    \LineComment[1]{Possibly adds a peer's block to the DAG}
    \If{$parents(b)\subset dag$}
        \State $q \gets \lbrace b \rbrace$
        \For {$b$ from $q$}
            \State $dag\gets dag + b$
            \ParComment{$satisfied$ returns the set of pending blocks with all parents now in $dag$}
            \State $s \gets satisfied(dag, pending)$
            \State $q \gets q + s$
            \State $pending \gets pending - s$
        \EndFor
    \Else
        \State $pending\gets pending + b$
    \EndIf
\EndProcedure \procSep
\Procedure{ReceiveEvent}{$event$}
    \LineComment[1]{Uses: $dag,\underline{waiting}$}
    \LineComment[1]{Possibly add an event to our waiting set}
    \If{$parents(event)\subset dag \cup waiting$}
        \State $waiting\gets waiting + event$
    \EndIf
\EndProcedure \procSep
\Procedure{TimeExpire}{}
    \LineComment[1]{Uses: $\underline{dag},\underline{waiting}$}
    \LineComment[1]{Create a new block at a regular interval}
    \State $b\gets new block$
    \State $b.parents\gets leaves(dag)$
    \State $b.transactions\gets waiting$
    \State $waiting\gets \varnothing$
    \State $dag\gets dag + b$
    \State broadcast $b$ to peers
\EndProcedure
\end{algorithmic}
\end{algorithm}

\begin{algorithm}[ht]
\caption{Casanova-\textit{conflict-attest}}\label{conflict-attest}
\begin{algorithmic}[1]
\State $dag \gets (\lbrace genesis \rbrace, \varnothing)$
\State $pending, waiting, inprogress, resolved \gets \varnothing$
\ParComment{Symbols underlined in the "Uses:" comments may be modified in the corresponding procedure}
\Procedure{ReceiveBlock}{$b$}
    \LineComment[1]{Uses: $\underline{dag}, \underline{pending}, \underline{inprogress}, resolved$}
    \LineComment[1]{Possibly adds a peer's block to the DAG}
    \If{$parents(b)\subset dag$}
        \State $q \gets \lbrace b \rbrace$
        \For{$b$ from $q$}
            \State $t\gets transactions(b)$
            \ParForAll{$e \in t$ s.t. $\exists e' \in dag \and e' \neq e \\ \and index(e') = index(e)$}
                \LineComment[1]{For every newly found conflict}
                \State $i \gets index(e)$
                \If{$i \in inprogress$}
                    \ParState{add $e$ to possible alternatives of ongoing consensus protocol for $i$}
                \ElsIf{$i \not\in resolved$}
                    \State $inprogress \gets inprogress + i$
                    \State begin side consensus protocol for $i$
                \EndIf
            \EndFor
            \State $dag \gets dag + b$
            \ParComment{$satisfied$ returns the set of pending blocks with all parents now in $dag$}
            \State $s \gets satisfied(dag, pending)$
            \State $q \gets q + s$
            \State $pending \gets pending - s$
        \EndFor
    \Else
        \State $pending\gets pending + b$
    \EndIf
\EndProcedure \procSep
\Procedure{ReceiveEvent}{$event$}
    \LineComment[1]{Uses: $dag,inprogress,resolved, \underline{waiting}$}
    \LineComment[1]{Possibly add an event to our waiting set}
    \If{$parents(event)\subset dag \cup waiting$}
        \State $i \gets index(event)$
        \If{$i \not\in inprogress \cup resolved \cup waiting$}
            \State $waiting\gets waiting + event$
        \EndIf
    \EndIf
\EndProcedure \procSep
\Procedure{SideConsensusAchieved}{$i,resolution$}
    \LineComment[1]{Uses: $\underline{inprogress}, \underline{resolved}, \underline{waiting}$}
    \LineComment[1]{Called when the side algorithm reaches a decision}
    \State $inprogress \gets inprogress - i$
    \State $resolved \gets resolved + i$
    \State $waiting \gets waiting + resolution$
\EndProcedure \procSep
\Procedure{TimeExpire}{}
    \State Unchanged from Casanova-attest
\EndProcedure
\end{algorithmic}
\end{algorithm}

\begin{algorithm}[ht]
\caption{Casanova-\textit{conflict-exclude}}\label{conflict-exclude}
\begin{algorithmic}[1]
\State $dag \gets (\lbrace genesis \rbrace, \varnothing)$
\State $pending, waiting, inprogress, resolved \gets \varnothing$
\ParComment{Symbols underlined in the "Uses:" comments may be modified in the corresponding procedure}
\Procedure{ReceiveBlock}{$b$}
    \LineComment[1]{Uses: $\underline{dag}, \underline{pending}, \underline{inprogress}, \underline{resolved}$}
    \LineComment[1]{Possibly adds a peer's block to the DAG}
    \If{$parents(b)\subset dag$}
        \State $q \gets \lbrace b \rbrace$
        \For {$b$ from $q$}
            \State $t\gets transactions(b)$
            \ParForAll{$e \in t$ s.t. $\exists e' \in dag \and e' \neq e \\ \and index(e') = index(e)$}
                \LineComment[1]{For every newly found conflict}
                \State $i \gets index(e)$
                \ParIf{$dag + b$ has an $FTM$-observed set for some $e \in \mathcal{E}_{i}$}
                    \State $resolved \gets resolved + i$
                    \State $excluded \gets true$
                \Else
                    \State $excluded \gets false$
                \EndIf
                \If{$i \in inprogress$ and $excluded$}
                    \ParState{stop participating in ongoing consensus protocol for $i$}
                    \State $inprogress \gets inprogress - i$
                \ElsIf{$i \in inprogress$ and $!excluded$}
                    \ParState{add $e$ to possible alternatives of ongoing consensus protocol for $i$}
                \ElsIf{$i \not\in resolved$ and $!excluded$}
                    \State $inprogress \gets inprogress + i$
                    \State begin side consensus protocol for $i$
                \EndIf
            \EndFor
            \State $dag\gets dag + b$
            \ParComment{$satisfied$ returns the set of pending blocks with all parents now in $dag$}
            \State $s \gets satisfied(dag, pending)$
            \State $q \gets q + s$
            \State $pending \gets pending - s$
        \EndFor
    \Else
        \State $pending\gets pending + b$
    \EndIf
\EndProcedure \procSep
\Procedure{ReceiveEvent}{$event$}
    \State Unchanged from Casanova-conflict-attest
\EndProcedure \procSep
\Procedure{SideConsensusAchieved}{$i,resolution$}
    \State Unchanged from Casanova-conflict-attest
\EndProcedure \procSep
\Procedure{TimeExpire}{}
    \State Unchanged from Casanova-conflict-attest
\EndProcedure
\end{algorithmic}
\end{algorithm}

\begin{algorithm}[ht]
\caption{Casanova}\label{casanova}
\begin{algorithmic}[1]
\State $dag \gets (\lbrace genesis \rbrace, \varnothing)$
\State $pending, waiting, inprogress, resolved \gets \varnothing$
\ParComment{Symbols underlined in the "Uses:" comments may be modified in the corresponding procedure}
\Procedure{ReceiveBlock}{$b$}
    \LineComment[1]{Uses: $\underline{dag}, \underline{pending}, \underline{inprogress}, \underline{resolved}$}
    \LineComment[1]{Possibly adds a peer's block to the DAG}
    \If{$parents(b)\subset dag$}
        \State $q \gets \lbrace b \rbrace$
        \For {$b$ from $q$}
            \State $t\gets transactions(b)$
            \ForAll{$vote \in votes(b)$}
                \State $i \gets index(vote)$
                \ParIf{$dag + b$ has a $k$-observed set for some $e_{i} \in \mathcal{E}_{i}$ and some round $r$}
                    \State $resolved \gets resolved + i$
                    \State $inprogress \gets inprogress - i$ \LineComment{$i$ may not belong to $inprogress$}
                \EndIf
            \EndFor
            \ParForAll{$e \in t$ s.t. $\exists e' \in dag \and e' \neq e \\ \and index(e') = index(e)$}
                \LineComment[1]{For every newly found conflict}
                \If{$i \in inprogress$}
                    \State add $e$ to list of alternatives for $\mathcal{E}_{i}$
                \ElsIf{$i \not\in resolved$}
                    \State $inprogress \gets inprogress + i$
                    \ParState{Set up rounds for $i$, and add $e, e'$ to list of alternatives for $\mathcal{E}_{i}$}
                \EndIf
            \EndFor
            \State $dag\gets dag + b$
            \ParComment{$satisfied$ returns the set of pending blocks with all parents now in $dag$}
            \State $s \gets satisfied(dag, pending)$
            \State $q \gets q + s$
            \State $pending \gets pending - s$
        \EndFor
    \Else
        \State $pending\gets pending + b$
    \EndIf
\EndProcedure
\algstore{casanova}
\end{algorithmic}
\end{algorithm}
\begin{algorithm}[ht]
\begin{algorithmic}[1]
\algrestore{casanova}
\Procedure{ReceiveEvent}{$event$}
    \State Unchanged from Casanova-conflict-exclude
\EndProcedure \procSep
\Function{ComputeVotes}{}
    \LineComment[1]{Uses: $dag,inprogress$}
    \State $votes \gets \varnothing$
    \ParFor{$i \in inprogress$ where $i$ starts a new round this block}
        \ParIf{$dag$ contains score of $FTM$ or greater in any round for $i$}
            \ParState{$votes \gets votes + $ vote for $e \in \mathcal{E}_{i}$ that achieved FTM in most recent round}
            \LineComment{The above 2 lines handle the locking logic}
        \Else
            \ParState{$votes \gets votes + $ vote for $e \in \mathcal{E}_{i}$ with lowest hash}
        \EndIf
    \EndFor
    \State \textbf{return} $votes$
\EndFunction \procSep
\Procedure{TimeExpire}{}
    \LineComment[1]{Uses: $\underline{dag}, \underline{waiting}$}
    \LineComment[1]{Create a new block at a regular interval}
    \State $b\gets new block$
    \State $b.parents\gets leaves(dag)$
    \State $b.transactions\gets waiting$
    \State $waiting \gets \varnothing$
    \State $b.votes\gets ComputeVotes(dag)$
    \State $dag\gets dag + b$
    \State broadcast $b$ to peers
\EndProcedure
\end{algorithmic}
\end{algorithm}

\clearpage
\section{Conclusion}
Casanova is an attestation-based Byzantine fault tolerant algorithm for ongoing consensus on a causally linked DAG of transactions. It is optimized for the case where the majority of transactions never see conflicts. It requires that transaction conflicts partition the set of possible transactions, a requirement that can be met in a ledger, but is not met \textit{e.g.} in Bitcoin\cite{BitcoinWhitePaper}{.} A programmable blockchain using this consensus algorithm would need to carefully match the programming model to this requirement. Overall we expect that the requirements can be met in a variety of useful applications, and that the algorithm will scale very well.

We encourage researchers to extend these ideas to optimize other kinds of consensus protocols. We warmly welcome efforts to implement Casanova and measure its performance.


\section{Acknowledgment}
Some of our initial ideas were inspired by Vlad Zamfir and his colleagues' work on Casper \cite{Casper}{.}
The integrated consensus protocol borrows heavily from Tendermint \cite{Tendermint} and Algorithm 2\textsuperscript{2} from DLS\cite{DLS}{.}


\bibliography{Bibliography.bib}{}

\begin{thebibliography}{1}
\providecommand{\url}[1]{#1}
\csname url@samestyle\endcsname
\providecommand{\newblock}{\relax}
\providecommand{\bibinfo}[2]{#2}
\providecommand{\BIBentrySTDinterwordspacing}{\spaceskip=0pt\relax}
\providecommand{\BIBentryALTinterwordstretchfactor}{4}
\providecommand{\BIBentryALTinterwordspacing}{\spaceskip=\fontdimen2\font plus
\BIBentryALTinterwordstretchfactor\fontdimen3\font minus
  \fontdimen4\font\relax}
\providecommand{\BIBforeignlanguage}[2]{{%
\expandafter\ifx\csname l@#1\endcsname\relax
\typeout{** WARNING: IEEEtran.bst: No hyphenation pattern has been}%
\typeout{** loaded for the language `#1'. Using the pattern for}%
\typeout{** the default language instead.}%
\else
\language=\csname l@#1\endcsname
\fi
#2}}
\providecommand{\BIBdecl}{\relax}
\BIBdecl

\bibitem{BitcoinWhitePaper}
\BIBentryALTinterwordspacing
S.~Nakamoto. (2009, October) Bitcoin: A peer-to-peer electronic cash system.
  [Online]. Available: \url{http://bitcoin.org/bitcoin.pdf}
\BIBentrySTDinterwordspacing

\bibitem{PaLa}
T.-H.~H. Chan, R.~Pass, and E.~Shi, ``Pala: A simple partially synchronous
  blockchain,'' \emph{IACR Cryptology ePrint Archive}, vol. 2018, p. 981, 2018.

\bibitem{FLP}
\BIBentryALTinterwordspacing
M.~J. Fischer, N.~A. Lynch, and M.~S. Paterson, ``Impossibility of distributed
  consensus with one faulty process,'' \emph{J. ACM}, vol.~32, no.~2, pp.
  374--382, Apr 1985. [Online]. Available:
  \url{http://doi.acm.org/10.1145/3149.214121}
\BIBentrySTDinterwordspacing

\bibitem{DLS}
\BIBentryALTinterwordspacing
C.~Dwork, N.~Lynch, and L.~Stockmeyer, ``Consensus in the presence of partial
  synchrony,'' \emph{J. ACM}, vol.~35, no.~2, pp. 288--323, Apr. 1988.
  [Online]. Available: \url{http://doi.acm.org/10.1145/42282.42283}
\BIBentrySTDinterwordspacing

\bibitem{Casper}
\BIBentryALTinterwordspacing
V.~Zamfir, N.~Rush, A.~Asgaonkar, and G.~Piliouras. (2018, Nov) Introducing the
  ``minimal cbc casper'' family of consensus protocols. [Online]. Available:
  \url{https://github.com/cbc-casper/cbc-casper-paper/blob/5f1dc67972180b90c7a411995082235bbfff2059/cbc-casper-paper-draft.pdf}
\BIBentrySTDinterwordspacing

\bibitem{Tendermint}
\BIBentryALTinterwordspacing
J.~K. Kwon. (2014) Tendermint : Consensus without mining. [Online]. Available:
  \url{https://cdn.relayto.com/media/files/LPgoWO18TCeMIggJVakt\_tendermint.pdf}
\BIBentrySTDinterwordspacing

\end{thebibliography}
\bibliographystyle{IEEEtran}

\appendix \label{Algorithm22}

{\bf Algorithm 2\textsuperscript{2} from DLS}

We provide a brief outline of Algorithm 2\textsuperscript{2} from ``Consensus in the Presence of Partial Synchrony.'' \cite{DLS}{.}
The algorithm is organized into \textit{phases} each phase consisting of four \textit{rounds}: \textit{accept, propose, acknowledge, lock-release}. Within a phase, each round is a fixed length. In each subsequent phase, the round length is increased by some constant $\delta$. The algorithm starts with phase 0, and each phase has an associated leader, in round robin fashion. Each message includes the phase to which it belongs, and is authenticated. Processors may have locks on a value, and each lock has an associated phase, and associated proof of acceptability. If a processor has a lock on a value, that is its only acceptable value. If it has a lock on multiple values, no value is acceptable. Otherwise, any \textit{proper} value is acceptable. We can ignore \textit{proper} for our purposes. 

In the \textit{accept} round, each processor sends an authenticated message containing a list of acceptable values to the leader of the phase.

In the \textit{propose} round, if the leader received $N - f$ messages stating that a value $v$ was acceptable it may propose $v$. If more than one such $v$ is acceptable, it chooses arbitrarily. It sends to each processor an authenticated message (lock v, proof). If it did not receive enough messages, it refrains from sending a lock message.

If a processor receives a valid lock message during the \textit{propose} round, it takes a lock on $v$ with associated phase $k$. If it already had a lock on $v$, it releases the earlier lock. It does not release locks on other values at this time. If it receives more than one such message, it takes locks on all such values.

In the \textit{acknowledge} round, processors send acknowledgments for the lock messages received in the \textit{propose} round to the round leader. If the leader receives an $NFM$ of acknowledgements for their lock message, they decide on $v$. The leader continues to participate.

In the \textit{lock release} round, processors broadcast the lock messages that caused them to lock a specific value for any locks they currently hold. If they receive a lock on a value $w$ with phase $k'$ and they have a lock on $v \neq w$ at phase $k' \geq k$, the processor releases its lock on $v$.

\end{document}